\documentclass[pra,aps,preprint,showpacs]{revtex4}
\usepackage{epsfig,amssymb,amsfonts}
\usepackage{graphicx}
\usepackage{dcolumn}
\usepackage{bm}

\def\opone{\leavevmode\hbox{\small1\kern-3.8pt\normalsize1}}

\begin{document}

\title{Entangled quantum heat engines based on two two-spin systems with Dzyaloshinski-Moriya anisotropic antisymmetric interaction}
\author{Guo-Feng Zhang\footnote{Corresponding author.}\footnote{Email:
gf1978zhang@buaa.edu.cn}} \affiliation{Department of physics, School
of sciences, Beijing University of Aeronautics and Astronautics,
Xueyuan Road No. 37, Beijing 100083, People's Republic of China}

\begin{abstract}
We construct an entangled quantum heat engine (EQHE) based on two
two-spin systems with Dzyaloshinski-Moriya (DM) anisotropic
antisymmetric interaction. By applying the explanations of heat
transferred and work performed at the quantum level in Kieu¡¯s work
[PRL, 93, 140403 (2004)], the basic thermodynamic quantities, i.e.,
heat transferred, net work done in a cycle and efficiency of EQHE
are investigated in terms of DM interaction and concurrence. The
validity of the second law of thermodynamics is confirmed in the
entangled system. It is found that there is a same efficiency for
both antiferromagnetic and ferromagnetic cases, and the efficiency
can be controlled in two manners: 1. only by spin-spin interaction
$J$ and DM interaction $D$; 2. only by the temperature $T$ and
concurrence $C$. In order to obtain a positive net work, we need not
entangle all qubits in two two-spin systems and we only require the
entanglement between qubits in a two-spin system not be zero. As the
ratio of entanglement between qubits in two two-spin systems
increases, the efficiency will approach infinitely the classical
Carnot one. An interesting phenomenon is an abrupt transition of the
efficiency when the entanglements between qubits in two two-spin
systems are equal.
\end{abstract}


\maketitle

\section{Introduction}
The dominating feature of an industrial society is its ability to
utilize, whether for wise or unwise ends, sources of energy other
than the muscles of men or animals. Except for waterpower, where
mechanical energy is directly available, most energy supplies are in
the form of potential energy of molecular or nuclear aggregations.
In chemical or nuclear reactions, some of this potential energy is
released and converted to random molecular kinetic energy. Heat can
be withdrawn and utilized for heating buildings, for cooking, or for
maintaining a furnace at a high temperature in order to carry out
other chemical or physical processes. But to operate a machine, one
of the problems of the mechanical engineers is to withdraw heat from
a high-temperature source and convert as large as a fraction as
possible to mechanical energy. We can solve the problems by using
heat engines which can  extract energy from its environment in the
form of heat and do useful work. We consider for simplicity a heat
engine in which the so-called ``working substance'' is carried
through a cycle process, that is, a sequence of processes in which
it eventually returns to its original state. All the heat engines
absorb heat from a source at a high temperature, perform some
mechanical work, and reject heat at a lower temperature. Consider a
heat engine operating in a cycle over and over again and let $Q_{h}$
and $Q_{l}$ stand for the heats absorbed and rejected by the working
substance per cycle. The net heat absorbed is $Q=Q_{h}+Q_{l}$. The
useful output of the engine is the net work $W$ done by the working
substance, and from the first law: $W=Q=Q_{h}+Q_{l}$. The heat
absorbed is usually obtained from the combustion fuel. The heat
rejected ordinarily has no economic value. The thermal efficiency of
a cycle is defined as the ratio of the useful work to the heat
absorbed: $\eta_{T}=W/Q_{h}=(Q_{h}+Q_{l})/Q_{h}$. Because of
friction losses, the useful work delivered by an engine is less than
the work $W$, and the overall efficiency is less than the thermal
efficiency. A Carnot cycle, in which all the heat input is supplied
at a single high temperature and all the heat output is rejected at
a single lower temperature, has an efficiency less than or equal to
the Carnot efficiency $\eta_{c}=(Q_{h}+Q_{l})/Q_{h}=1-T_{l}/T_{h}$,
where $T_{h}$ and $T_{l}$ are the temperature of the
high-temperature energy source and the low-temperature energy sink,
respectively. This is supported by the second law of thermodynamics
and numerous experimental evidences.

Quantum heat engines (QHEs), in contrast, operated by passing
quantum matter through a closed series of quantum adiabatic
processes and energy exchanges with heat baths, respectively, at
different stages of a cycle. Kieu\cite{T. D. Kieu} constructed a
quantum heat engine (QHE) which is a two-level quantum system and
undergoes quantum adiabatic process and energy exchanges with heat
baths at different stages in a work cycle. Armed with this class of
QHE and the interpretations of heat transferred and work performed
at the quantum level, he clarified some important aspects of the
second law of thermodynamics. Recently, the physics of
semiconductors with a spin-orbit interaction has attracted a lot of
attention, as it plays an important role for the emerging field of
semiconductor spintronics \cite{I. Zutic}. Sun et al. investigate
the equilibrium property of a mesoscopic ring with a spin-orbit
interaction, persistent spin current is studied \cite{sqf}.

However, none of the QHEs mentioned above involve the most
extraordinary phenomenon in quantum mechanics, i.e., quantum
entanglement \cite{aei,esc,pre,hor,chb}. To enrich research in QHEs,
an entangled QHE and the investigation about the influence of
entanglement on its thermodynamic characteristics should be
considered. The entanglement of quantum spin systems at finite
temperatures has been extensively studied for various
interaction styles\cite{mca,dgu,xw,mas}. Recently, Zhang et al.\cite{T. Zhang}
extended Kieu's work by considering the quantum engine with a
two-qubit (isotropic) Heisenberg XXX spin system as the working substance.
The spin is subject to a constant external magnetic field. The
purpose of their paper is to analyze the effect of quantum
entanglement on the efficiency of the quantum engine, but only a
constant external magnetic field and entanglement was considered in
that study. Also, they only investigated the antiferromagnetic
coupling case, that is incomplete. For a ferromagnetic coupling spin
chain, whether quantum heat engine is available or not deserve
investigation. Moreover in the theoretical analysis we think it is
very interesting to investigate the effects of spin-orbit coupling
on the basic thermodynamic quantities and should be included. In
other words, quantum heat engine is a new physics concept,
spin-orbit coupling in spin system plays an important role in the
preparation of quantum apparatus. The effect of spin-orbit coupling
on quantum heat engine is a very interesting topic and deserves
studying. In this paper, We construct an entangled quantum heat engine (EQHE) based on two
two-spin systems with a kind of spin-orbit coupling interaction.
The positive net work condition in terms of spin-orbit interaction and concurrence will be given. The
validity of the second law of thermodynamics at the quantum level is
again confirmed.

\section{Entangled quantum heat engine based on two two-spin systems with DM interaction}
Our EQHE is based on the Heisenberg model with DM interaction, which
can be described by
\begin{equation}
\label{1}
H_{DM}=\frac{J}{2}[(\sigma_{1x}\sigma_{2x}+\sigma_{1y}\sigma_{2y})+\overrightarrow{D}\cdot(\overrightarrow{\sigma_{1}}\times\overrightarrow{\sigma_{2}})],
\end{equation}
here $J$ is the real coupling coefficient and $\overrightarrow{D}$
is the DM vector coupling. The DM anisotropic antisymmetric
interaction arises from spin-orbit coupling \cite{I. Dzyaloshinskii,T. Moriya}. The
coupling constant $J>0$ corresponds to the antiferromagnetic case
and $J<0$ to the ferromagnetic case. For simplicity, we choose
$\overrightarrow{D}=D\overrightarrow{z}$, then the Hamiltonian
$H_{DM}$ becomes
\begin{eqnarray}
\label{2}
H_{DM}&=&\frac{J}{2}[\sigma_{1x}\sigma_{2x}+\sigma_{1y}\sigma_{2y}+D(\sigma_{1x}\sigma_{2y}-\sigma_{1y}\sigma_{2x})]
\nonumber
\\&=&J[(1+iD)\sigma_{1+}\sigma_{2-}+(1-iD)\sigma_{1-}\sigma_{2+}].
\end{eqnarray}
Without loose of generality, we define $|0\rangle$ $(|1\rangle)$ as
the ground (excited) state of a two-level particle. The eigenvalues
and eigenvectors of $H_{DM}$ are given by
\begin{eqnarray}
|\Psi^{1}\rangle=|00\rangle, |\Psi^{2}\rangle=|11\rangle,
|\Psi^{3}\rangle=\frac{1}{\sqrt{2}}\left\{|01\rangle+e^{i\theta}|10\rangle\right\},
|\Psi^{4}\rangle=\frac{1}{\sqrt{2}}\left\{|01\rangle-e^{i\theta}|10\rangle\right\},
\end{eqnarray}
with $E_{1}=E_{2}=0$, $E_{3}=J\sqrt{1+D^{2}}=-E_{4}$ and
$\theta=\arctan D$. As the thermal fluctuation (temperature
$\emph{T}$) is introducing into the system, the state of a typical
solid-state system at thermal equilibrium (temperature $T$) is
$\rho(T)=\frac{1}{Z}e^{-\beta\emph{H}_{DM}}$, where $\emph{H}_{DM}$
is the Hamiltonian, $Z=tre^{-\beta\emph{H}_{DM}}$ is the partition
function, $\beta=1/(kT)$ and $k$ is Boltzmann's constant, for
simplicity, we write $k=1$. The entanglement between two qubits in
this model is known as \cite{X. G. W}:
\begin{eqnarray}
C=\left\{%
\begin{array}{ll}
     \frac{\sinh(|J|\sqrt{1+D^{2}}/T)-1}{\cosh(|J|\sqrt{1+D^{2}}/T)+1}, & \hbox{if $|J|\sqrt{1+D^{2}}>T\texttt{arcsinh(1)}$;} \\
    0, & \hbox{if $|J|\sqrt{1+D^{2}}\leq T\texttt{arcsinh(1)}$.} \\
\end{array}%
\right.
\end{eqnarray}
Here, we use concurrence directly as the measurement of
entanglement, since there is a one-to-one correspondence between the
entanglement of formation and concurrence\cite{hil,woo}. The
concurrence $C=0$ indicates the vanishing entanglement. The critical
temperature
$T_{c}=\frac{|J|\sqrt{1+D^{2}/T}}{\texttt{arcsinh(1)}}\approx1.1346|J|\sqrt{1+D^{2}}$
above which the concurrence is zero.

Based on the Kieu¡¯s explanations of heat transferred and work
performed at the quantum level \cite{T. D. Kieu}: The expectation
value of the measured energy of a quantum system with discrete
energy levels is: $U=<E>=\sum_{i}p_{i}E_{i}$, where $E_{i}$ are the
energy levels and $p_{i}$ are the corresponding occupation
probabilities. Infinitesimally,
$dU=\sum_{i}\left\{p_{i}dE_{i}+E_{i}dp_{i}\right\}$, from which we
can obtain the following identifications for infinitesimal heat
transferred $\bar{d}Q$ and work done $\bar{d}W$
\begin{equation}
\bar{d}Q:=\sum_{i}E_{i}dp_{i}, \bar{d}W:=\sum_{i}p_{i}dE_{i}.
\end{equation}
Mathematically speaking, these are not total differentials but are
path dependent. These expressions interpret heat as a change in the
occupation probabilities but not in the distribution of the energy
eigenvalues themselves; and work done as the redistribution of the
energy eigenvalues but not of the occupation probabilities of each
energy level \cite{E. Sch}. Thus, $dU=\bar{d}Q+\bar{d}W$, which is
an expression of the first law of thermodynamics.

To set the stage, we describe the four quantum thermodynamic
processes of the quantum cycle. In the following, we consider as
working substance a bipartite quantum system consisting of two
subsystems, $A$ and $B$, with the above Hamiltonian (2). A cycle of the quantum heat engine consists of four stages:

$\langle1\rangle$ The system has the probability $p_{i0}$ (i
=1,2,3,4) to be in each of its four eigenstates, respectively. By
contacting with a heat bath at temperature $T_{h}$ for some time,
the occupation probability of each eigenstate becomes $p_{i1}$. The
quantum state of the system is given by the density operator
$\rho^{1}_{AB}=\sum_{i}p_{i1}|\Psi^{i1}\rangle\langle\Psi^{i1}|,
(i=1,2,3,4)$ with $p_{i1}=\exp(-E_{i1}/kT_{h})/Z_{1}$ and
$Z_{1}=\sum_{i}\exp(-E_{i1}/kT_{h})$. In other words, we assume the
system is initially in thermal equilibrium with a heat bath at
temperature $T_{h}$. In our model, we have $E_{i1}=E_{i}$,
$|\Psi^{i1}\rangle=|\Psi^{i}\rangle$ with $J=J_{1}$ and $D=D_{1}$.
Only heat is transferred in this process due to the change in the
occupation probability. $\langle2\rangle$ The system is then
isolated from the heat bath and undergoes a quantum  adiabatic
process, with $J$ changing from $J_{1}$ to $J_{2}$ and $D$ changing
from $D_{1}$ to $D_{2}$. Provided the rate of change is sufficiently
slow, $p_{i1}$'s are maintained throughout according to the quantum
adiabatic theorem. At the end of process $\langle2\rangle$, the
system has the probability $p_{i1}$ in the eigenstate
$|\Psi^{i1}\rangle$. An amount of work is performed by the system,
but no heat is transferred. $\langle3\rangle$ The system is next
brought into some kind of contact with another heat bath at
temperature $T_{l}$$(T_{l}<T_{h})$ for some time. After the
irreversible thermalization process, the quantum state of the system
is given by the density operator
$\rho^{2}_{AB}=\sum_{i}p_{i2}|\Psi^{i2}\rangle\langle\Psi^{i2}|,
(i=1,2,3,4)$ with $p_{i2}=\exp(-E_{i2}/kT_{l})/Z_{2}$ and
$Z_{2}=\sum_{i}\exp(-E_{i2}/kT_{l})$. Here, we have $E_{i2}=E_{i}$,
$|\Psi^{i2}\rangle=|\Psi^{i}\rangle$ with $J=J_{2}$ and $D=D_{2}$.
Only heat is transferred in this process to yield a change in the
occupation probabilities, and the heat transferred is given by
$Q_{l}=\sum_{i}E_{i2}(p_{i2}-p_{i1})$. $\langle4\rangle$ The system
is removed from the heat bath and undergoes a quantum adiabatic
process, with for instance $J$ changing from $J_{2}$ to $J_{1}$ and
$D$ changing from $D_{2}$ to $D_{1}$. And the occupation probability
of each eigenstate is maintained, that is, $p_{i2}=p_{i0}$. An
amount of work is performed on the system, but no heat is
transferred during the process. So we can know the heat transferred
in the first process is
$Q_{h}=\sum_{i}E_{i1}(p_{i1}-p_{i0})=\sum_{i}E_{i1}(p_{i1}-p_{i2})$.
From the law of conservation of energy, the net work done by the
EQHE in two quantum adiabatic processes is given by $W=
Q_{h}+Q_{l}=\sum_{i}(E_{i1}-E_{i1})(p_{i1}-p_{i2}),(i=1,2,3,4)$. For
a actual engine,  $W>0$ is required, it means that three possible
cases, i.e., $Q_{h}>-Q_{l}>0$, or $Q_{l}>-Q_{h}>0$, or $Q_{h}>0$ and
$Q_{l}>0$.

\section{Basic thermodynamic quantities in terms of DM interaction}

After a simple calculation, we can get the heat absorbed $Q_{h}$,
the heat refused $Q_{l}$ and the net work
\begin{eqnarray}
Q_{h}&=&\sqrt{1+D^{2}_{1}}J_{1}(\tanh[\frac{J_{2}\sqrt{1+D_{2}^{2}}}{2T_{l}}]-\tanh[\frac{J_{1}\sqrt{1+D_{1}^{2}}}{2T_{h}}]),\nonumber
\\Q_{l}&=&\sqrt{1+D^{2}_{2}}J_{2}(\tanh[\frac{J_{1}\sqrt{1+D_{1}^{2}}}{2T_{h}}]-\tanh[\frac{J_{2}\sqrt{1+D_{2}^{2}}}{2T_{l}}]),\nonumber
\\W&=&(J_{2}\sqrt{1+D_{2}^{2}}-J_{1}\sqrt{1+D_{1}^{2}})(\tanh[\frac{J_{1}\sqrt{1+D_{1}^{2}}}{2T_{h}}]-\tanh[\frac{J_{2}\sqrt{1+D_{2}^{2}}}{2T_{l}}]).
\end{eqnarray}
Thus, under the postulate $T_{l}<T_{h}$, in order to make $W>0$, we
must have
\begin{equation}
\frac{J_{1}\sqrt{1+D_{1}^{2}}}{T_{h}}<\frac{J_{2}\sqrt{1+D_{2}^{2}}}{T_{l}}\Leftrightarrow
T_{h}>T_{l}\left(\frac{J_{1}\sqrt{1+D_{1}^{2}}}{J_{2}\sqrt{1+D_{2}^{2}}}\right).
\end{equation}
for $J_{2}\sqrt{1+D_{2}^{2}}<J_{1}\sqrt{1+D_{1}^{2}}$. If the DM
interaction is turned off, Eq.(7) will be replaced by
$T_{h}>T_{l}\left(\frac{J_{1}}{J_{2}}\right)$. These conditions are
rigorous in contradistinction to the classical requirement that
$T_{h}$ only needs to be larger than $T_{l}$.

The thermal efficiency of a cycle is defined as the ratio of the
useful work to the heat absorbed:
$\eta_{T}=W/Q_{h}=(Q_{h}+Q_{l})/Q_{h}$. After a straight
calculation, the efficiency can be given by
\begin{equation}
\eta=1-\frac{J_{2}\sqrt{1+D^{2}_{2}}}{J_{1}\sqrt{1+D^{2}_{1}}}.
\end{equation}
We can find that the efficiency is independent of temperature and
can be controlled only by spin-spin interaction $J$ and DM
interaction $D$. Although this is true, in order to obtain a
positive net work, the requirement about temperature Eq.(7) must be
satisfied.

\section{The effect of concurrence on basic thermodynamic quantities}

The entanglement under our consideration is that of two thermal
equilibrium states at the end of stage $\langle1\rangle$ and stage
$\langle3\rangle$, denoted by $C_{1}$ and $C_{2}$, respectively.
They are
\begin{eqnarray}
C_{1}=\left\{%
\begin{array}{ll}
     \frac{\sinh(|J_{1}|\sqrt{1+D_{1}^{2}}/T_{h})-1}{\cosh(|J_{1}|\sqrt{1+D_{1}^{2}}/T_{h})+1}    & \hbox{if $|J_{1}|\sqrt{1+D_{1}^{2}}>T_{h}\texttt{arcsinh(1)}$;} \\
    0                        & \hbox{if $|J_{1}|\sqrt{1+D_{1}^{2}}\leq T_{h}\texttt{arcsinh(1)}$.} \\
\end{array}%
\right.\nonumber
\end{eqnarray}
\begin{eqnarray}
C_{2}=\left\{%
\begin{array}{ll}
     \frac{\sinh(|J_{2}|\sqrt{1+D_{2}^{2}}/T_{l})-1}{\cosh(|J_{2}|\sqrt{1+D_{2}^{2}}/T_{l})+1}    & \hbox{if $|J_{2}|\sqrt{1+D_{2}^{2}}>T_{l}\texttt{arcsinh(1)}$;} \\
    0& \hbox{if $|J_{2}|\sqrt{1+D_{2}^{2}}\leq T_{l}\texttt{arcsinh(1)}$.} \\
\end{array}%
\right.
\end{eqnarray}
Here we will explore the relation between entanglement and basic
thermodynamics quantities.

Case1: The two-spin system is antiferromagnetic coupling, i.e.,
$J>0$. From Eq.(9), we find
\begin{eqnarray}
J_{1}&=&\frac{T_{l}}{\sqrt{1+D^{2}_{1}}}\ln\left(\frac{-1-C_{1}-\sqrt{2(1+C_{1})}}{C_{1}-1}\right),\nonumber
\\J_{2}&=&\frac{T_{h}}{\sqrt{1+D^{2}_{2}}}\ln\left(\frac{-1-C_{2}-\sqrt{2(1+C_{2})}}{C_{2}-1}\right).
\end{eqnarray}
By simple deduction,  we can obtain
\begin{eqnarray}
 Q_{h}&=&\sqrt{2}\ln[\frac{1}{-1+\sqrt{\frac{2}{1+C_{1}}}}](\sqrt{1+C_{2}}-\sqrt{1+C_{1}})T_{h}, \nonumber
\\Q_{l}&=&\sqrt{2}\ln[\frac{1}{-1+\sqrt{\frac{2}{1+C_{2}}}}](\sqrt{1+C_{1}}-\sqrt{1+C_{2}})T_{l}, \nonumber
\\W&=&-\sqrt{2}(\sqrt{1+C_{1}}-\sqrt{1+C_{2}})(\ln[\frac{1}{-1+\sqrt{\frac{2}{1+C_{1}}}}]T_{h}-\ln[\frac{1}{-1+\sqrt{\frac{2}{1+C_{2}}}}]T_{l}).
\end{eqnarray}
Therefore the positive net work condition indicates two
possible situations, i.e.,
\begin{equation}
\sqrt{1+C_{1}}<\sqrt{1+C_{2}}\quad\quad\texttt{and}\quad\quad\ln[\frac{1}{-1+\sqrt{\frac{2}{1+C_{1}}}}]T_{h}>\ln[\frac{1}{-1+\sqrt{\frac{2}{1+C_{2}}}}]T_{l},
\end{equation}
or
\begin{equation}
\sqrt{1+C_{1}}>\sqrt{1+C_{2}}\quad\quad\texttt{and}\quad\quad\ln[\frac{1}{-1+\sqrt{\frac{2}{1+C_{1}}}}]T_{h}<\ln[\frac{1}{-1+\sqrt{\frac{2}{1+C_{2}}}}]T_{l}.
\end{equation}
Equation (12) can be easily proved incompatible with $T_{h}>T_{l}$,
that is, case $Q_{h}<0$ and $Q_{l}>0$ is impossible. So far we have
clarified case $Q_{h}>-Q_{l}>0$ is the only possible case.  It
is shown that the second law of thermodynamics is valid in this case.
\begin{figure}
\begin{center}
\epsfig{figure=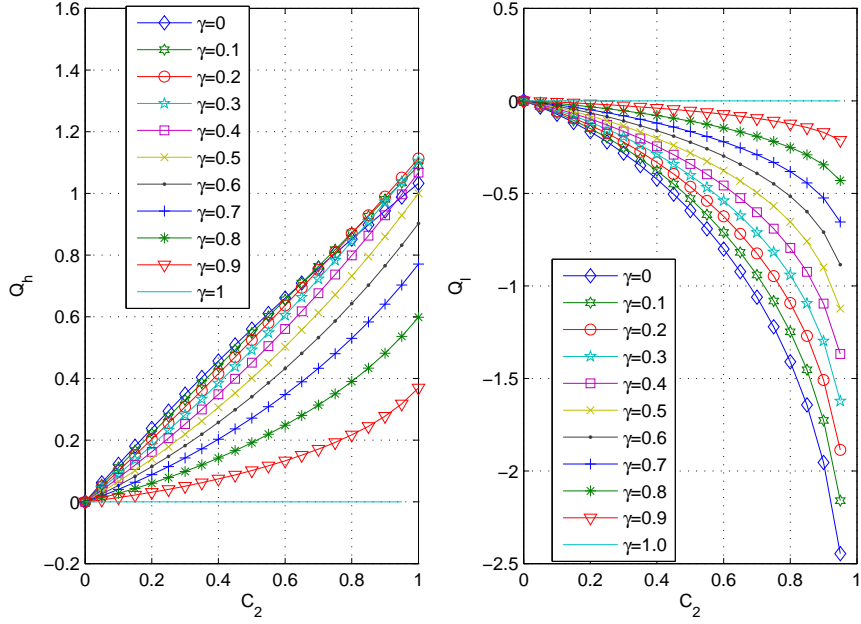,height=0.50\textwidth}
\end{center}
\caption{(Color online) The heat absorbed $Q_{h}$ and the one rejected
$Q_{l}$ are plotted vs the concurrence $C_{2}$ for different
$\gamma=C_{1}/C_{2}$ and $kT_{h}=2$, $kT_{l}=1$.}
\end{figure}
\begin{figure}
\begin{center}
\epsfig{figure=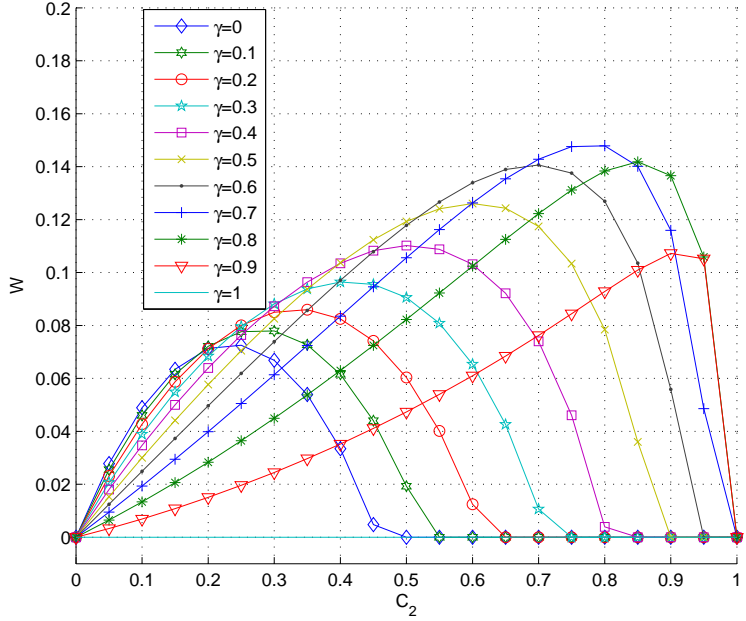,height=0.50\textwidth}
\end{center}
\caption{(Color online) The net work done by the EQHE is plotted vs the concurrence $C_{2}$ for different $\gamma=C_{1}/C_{2}$ and $kT_{h}=2$, $kT_{l}=1$.}
\end{figure}

Case2: The two-spin system is ferromagnetic coupling, i.e., $J<0$.
From Eq.(9), we find
\begin{eqnarray}
J_{1}&=&\frac{T_{l}}{\sqrt{1+D^{2}_{1}}}\ln\left(\frac{-1-C_{1}+\sqrt{2(1+C_{1})}}{C_{1}+1}\right),\nonumber
\\J_{2}&=&\frac{T_{h}}{\sqrt{1+D^{2}_{2}}}\ln\left(\frac{-1-C_{2}+\sqrt{2(1+C_{2})}}{C_{2}+1}\right).
\end{eqnarray}
Similarly,  we can obtain
\begin{eqnarray}
 Q_{h}&=&\sqrt{2}\ln[-1+\sqrt{\frac{2}{1+C_{1}}}](\sqrt{1+C_{1}}-\sqrt{1+C_{2}})T_{h}, \nonumber
\\Q_{l}&=&\sqrt{2}\ln[-1+\sqrt{\frac{2}{1+C_{2}}}](\sqrt{1+C_{2}}-\sqrt{1+C_{1}})T_{l}, \nonumber
\\W&=&\sqrt{2}(\sqrt{1+C_{1}}-\sqrt{1+C_{2}})(\ln[-1+\sqrt{\frac{2}{1+C_{1}}}]T_{h}-\ln[-1+\sqrt{\frac{2}{1+C_{2}}}]T_{l}).
\end{eqnarray}
In order to make $W>0$, we must have
\begin{equation}
\sqrt{1+C_{1}}<\sqrt{1+C_{2}}\quad\quad\texttt{and}\quad\quad\ln[-1+\sqrt{\frac{2}{1+C_{1}}}]T_{h}<\ln[-1+\sqrt{\frac{2}{1+C_{2}}}]T_{l},
\end{equation}
or
\begin{equation}
\sqrt{1+C_{1}}>\sqrt{1+C_{2}}\quad\quad\texttt{and}\quad\quad\ln[-1+\sqrt{\frac{2}{1+C_{1}}}]T_{h}>\ln[-1+\sqrt{\frac{2}{1+C_{2}}}]T_{l}.
\end{equation}
It is obvious that Eq.(16) is impossible for $T_{h}>T_{l}$. In fact, we can see that Eq.(11)(Eq.(12)) is the same with Eq.(15)(Eq.(16)).
Thus, we can conclude that the second law of thermodynamics is also valid in this case.

Variations of $Q_{h}$, $Q_{l}$ and the net work $W$ with $C_{2}$ are
given in fig.1 and fig.2 for antiferromagnetic coupling. Where we
define $\gamma =C_{1}/C_{2}$ as a measurement of the difference of
$C_{1}$ and $C_{2}$. From fig.1 and fig.2, We can see the engine
undergoes a trivial cycle since $Q_{h}=Q_{l}=W=0$ for $\gamma=1$.
That means, in order to make the engine be useful, the entanglement
between qubits for the two two-spin systems can not be equal. This
result accords with Eq.(11) and Eq.(15). For $\gamma=0$, i.e.,
$C_{1}=0$, the engine is available since $W>0$ for a range of
$C_{2}$. So long as $\gamma\neq1$, the range of $C_{2}$ in which
$W>0$ will be broaden as $\gamma$ increases. The results show that
we need not entangle qubits in two-spin system 1 and we only require
the entanglement in two-spin system 2 not be zero. The entanglement
in two-spin system 1 will enhance the validity of the engine.
\begin{figure}
\begin{center}
\epsfig{figure=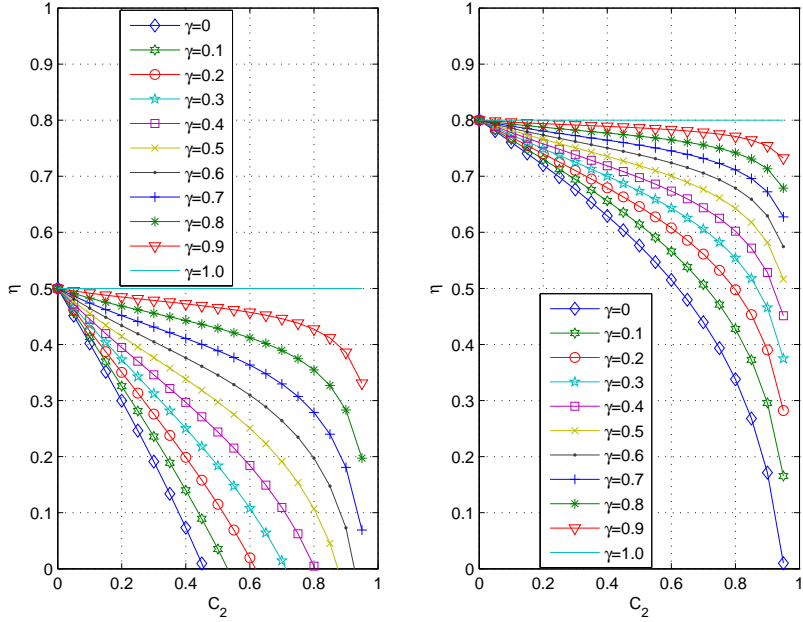,height=0.50\textwidth}
\end{center}
\caption{(Color online) The efficiency is plotted vs the concurrence
$C_{2}$ for different $\gamma=C_{1}/C_{2}$. And the left panel for
$kT_{h}/kT_{l}=2$, the right panel for $kT_{h}/kT_{l}=5$.}
\end{figure}

The efficiency in terms of concurrence can be given by
\begin{equation}
\eta(J>0)=\eta(J<0)=1-\frac{\ln[\frac{1}{-1+\sqrt{\frac{2}{1+C_{_{2}}}}}]T_{l}}{\ln[\frac{1}{-1+\sqrt{\frac{2}{1+C_{_{1}}}}}]T_{h}},\quad\quad\texttt{for} \sqrt{1+C_{1}}<\sqrt{1+C_{2}}.
\end{equation}
When $C_{2}=C_{1}$, $\eta(J>0)=\eta(J<0)=0$ since $Q_{h}=Q_{l}=W=0$
from Eqs.(11) and (15). In order to intuitively investigate how
entanglement affects the efficiency, we sketch the variation of
$\eta$ with $C_{2}$ for different ratio of $C_{1}$ to $C_{2}$ in
Fig.3. It is known that for a Carnot cycle, in which all the heat
input is supplied at a single high temperature and all the heat
output is rejected at a single lower temperature, has an efficiency
less than or equal to the Carnot efficiency
$\eta_{c}=(Q_{h}+Q_{l})/Q_{h}=1-T_{l}/T_{h}$. Here $\eta_{c}$ is
never achievable since $C_{1}$ can not be equal to $C_{2}$. From
fig.3, $\eta$ increases with $\gamma$ when $C_{2}$ is fixed and it
will approach $\eta_{c}$ infinitely. We can also see that $\eta$
decreases with $C_{2}$ and whether it will arrive at zero depends on
$\gamma$. The smaller $\gamma$ is, the bigger zero probability
$\eta$ has. When $C_{2}=C_{1}$, i.e., $\gamma=1$, we can find
$\eta=\eta_{c}$ from fig.3, but $Q_{h}=Q_{l}=W=0$ and $\eta$ should
be zero. So we say an abrupt transition of the efficiency occurs at
$\gamma=1$. Compared the left panel and the right one, $\gamma=0$,
i.e., $C_{1}=0$, $\eta$ can not be zero for a range of $C_{2}$ and
the range will be enlarged as the ratio $kT_{h}/kT_{l}$ increases.
For a same $\gamma$, $kT_{h}/kT_{l}$ must be greater in order to get
a higher efficiency. This result is consistent with the second law
of thermodynamics. In other words, the second law of thermodynamics
is proved to hold all the time even when entanglement is indeed
involved.

\section{Conclusions}
In conclusion, by quoting the quantum interpretations of heat and
work from Ref.\cite{T. D. Kieu}, we construct an entangled quantum
heat engine (EQHE) based on two two-spin systems with
Dzyaloshinski-Moriya (DM) anisotropic antisymmetric interaction. The
basic thermodynamic quantities, i.e., the heat transferred and the
work done in a cycle, and the efficiency of EQHE are investigated in
terms of DM interaction and concurrence. The condition for a
positive work is given. Four features of main results can be found
in this paper. First, the efficiency can be controlled only by
spin-spin interaction and DM interaction, it is independent of the
temperature. Second, we need not entangle qubits in two-spin system
1 and we only require the entanglement in two-spin system 2 not be
zero in order to obtain a positive net work. The entanglement in
two-spin system 1 will enhance the validity of the engine. Third,
for the same $\gamma$, the ratio $kT_{h}/kT_{l}$ must be greater in
order to get a higher efficiency. Fourth, an interesting phenomenon
is an abrupt transition of the efficiency when $C_{1}=C_{2}$. These
results in the paper are consistent with the second law of
thermodynamics. In other words, the second law of thermodynamics is
shown to be valid in this entangled system.

\section{acknowledgements}
It is a pleasure to thank the reviewer and editor for their many
fruitful discussions about the topic. This work was supported by the
National Natural Science Foundation of China (Grant No. 10604053)
and Beihang Lantian Project.

\end{document}